\documentclass[
reprint,
superscriptaddress,
amsmath,amssymb,
aps,
floatfix,
]{revtex4-2}

\usepackage{float}
\usepackage[utf8]{inputenc}
\usepackage{graphicx,import}
\graphicspath{{Figures/}}
\usepackage{bm}
\usepackage[colorlinks]{hyperref}
\usepackage[all]{hypcap}
\usepackage{xcolor}
\usepackage{xfrac}
\usepackage{braket}
\usepackage{siunitx}
\usepackage{changes}
\usepackage{lipsum}
\usepackage{layouts}	
\usepackage{amsmath}
\usepackage{upgreek}

\definecolor{bluegray}{RGB}{40,180,160}
\definecolor{navygray}{RGB}{110,140,170}
\definecolor{meadowgreen}{RGB}{0,128,0}
\definecolor{coolbrown}{RGB} {165,42,42}

\DeclareSIUnit{\sq}{\Box}

\newcommand{\secref}[1]{\hyperref[#1]{{Section~\ref{#1}}}}
\newcommand{\chapref}[1]{\hyperref[#1]{{Chapter~\ref{#1}}}}
\newcommand{\suppref}[1]{\hyperref[#1]{{App.~\ref{#1}}}}
\newcommand{\figref}[1]{\hyperref[#1]{{Fig.~\ref*{#1}}}}
\newcommand{\Figref}[1]{\hyperref[#1]{{Figure~\ref*{#1}}}}
\newcommand{\figrefadd}[2]{\hyperref[#1]{{Fig.~\ref*{#1}#2}}}
\newcommand{\Figrefadd}[2]{\hyperref[#1]{{Figure~\ref*{#1}#2}}}
\newcommand{\tabref}[1]{\hyperref[#1]{Tab.~\ref*{#1}}}
\newcommand{\refref}[1]{\hyperref[#1]{{Ref.~\ref*{#1}}}}
\renewcommand{\eqref}[1]{\hyperref[#1]{{Eq.~(\ref*{#1})}}}

\usepackage{listings}
\definecolor{codegreen}{rgb}{0,0.6,0}
\definecolor{codegray}{rgb}{0.5,0.5,0.5}
\definecolor{codepurple}{rgb}{0.58,0,0.82}
\definecolor{backcolour}{rgb}{0.95,0.95,0.92}

\lstdefinestyle{mystyle}{
  backgroundcolor=\color{backcolour}, commentstyle=\color{codegreen},
  keywordstyle=\color{magenta},
  numberstyle=\tiny\color{codegray},
  stringstyle=\color{codepurple},
  basicstyle=\ttfamily\footnotesize,
  breakatwhitespace=false,         
  breaklines=true,                 
  captionpos=t,                    
  keepspaces=true,                    
  showspaces=false,                
  showstringspaces=false,
  showtabs=false,                  
  tabsize=2
}

\newcommand{\revisex}[1]{{}}

\hypersetup{
citecolor={navygray}, 
linkcolor={navygray},
urlcolor={navygray},
}

\makeatletter
\newcommand*{\balancecolsandclearpage}{%
  \close@column@grid
  \cleardoublepage
  \twocolumngrid
}
\makeatother

\begin{document}

\title{Probing the memory of a superconducting qubit environment}

\author{Nicolas~Gosling}
\affiliation{IQMT,~Karlsruhe~Institute~of~Technology,~76131~Karlsruhe,~Germany}

\author{Denis~B\'en\^atre}
\thanks{First two authors contributed equally}
\affiliation{IQMT,~Karlsruhe~Institute~of~Technology,~76131~Karlsruhe,~Germany}

\author{Nicolas~Zapata}
\affiliation{IQMT,~Karlsruhe~Institute~of~Technology,~76131~Karlsruhe,~Germany}

\author{Paul~Kugler}
\affiliation{IQMT,~Karlsruhe~Institute~of~Technology,~76131~Karlsruhe,~Germany}

\author{Mitchell~Field}
\affiliation{IQMT,~Karlsruhe~Institute~of~Technology,~76131~Karlsruhe,~Germany}

\author{Sumeru~Hazra}
\affiliation{IQMT,~Karlsruhe~Institute~of~Technology,~76131~Karlsruhe,~Germany}

\author{Simon~Günzler}
\affiliation{IQMT,~Karlsruhe~Institute~of~Technology,~76131~Karlsruhe,~Germany}
\affiliation{PHI,~Karlsruhe~Institute~of~Technology,~76131~Karlsruhe,~Germany}

\author{Thomas~Reisinger}
\affiliation{IQMT,~Karlsruhe~Institute~of~Technology,~76131~Karlsruhe,~Germany}

\author{Martin~Spiecker}
\affiliation{IQMT,~Karlsruhe~Institute~of~Technology,~76131~Karlsruhe,~Germany}
\affiliation{PHI,~Karlsruhe~Institute~of~Technology,~76131~Karlsruhe,~Germany}

\author{Mathieu~F\'echant}
\affiliation{IQMT,~Karlsruhe~Institute~of~Technology,~76131~Karlsruhe,~Germany}

\author{Ioan~M.~Pop}
\email{ioan.pop@kit.edu}
\affiliation{IQMT,~Karlsruhe~Institute~of~Technology,~76131~Karlsruhe,~Germany}
\affiliation{PHI,~Karlsruhe~Institute~of~Technology,~76131~Karlsruhe,~Germany}
\affiliation{PI1,~Stuttgart~University,~70569~Stuttgart,~Germany}

\date{\today}

\begin{abstract}

Achieving fault tolerance with superconducting quantum processors requires qubits to operate within the regime of threshold theorems based on the Born-Markov approximation.
This approximation, which models dissipation as constant energy decay into a memoryless environment, breaks down when qubits couple to long-lived two-level systems (TLSs) that become polarized during operation and retain memory of past qubit states.
Here, we show that non-Poissonian quantum jump traces carry the information required to distinguish long-lived TLSs from the standard Markovian bath. 
By fitting the Solomon equations to measured quantum jumps dynamics arising naturally due to thermal fluctuations, we can disentangle the coupling of the qubit to the two environments.
Sweeping the qubit frequency reveals distinct peaks, each associated with a TLS that outlives the qubit, providing a handle to understand their microscopic origin.
\end{abstract}

\maketitle

\begin{figure*}[ht!]
\centering
\includegraphics[width=1\textwidth]{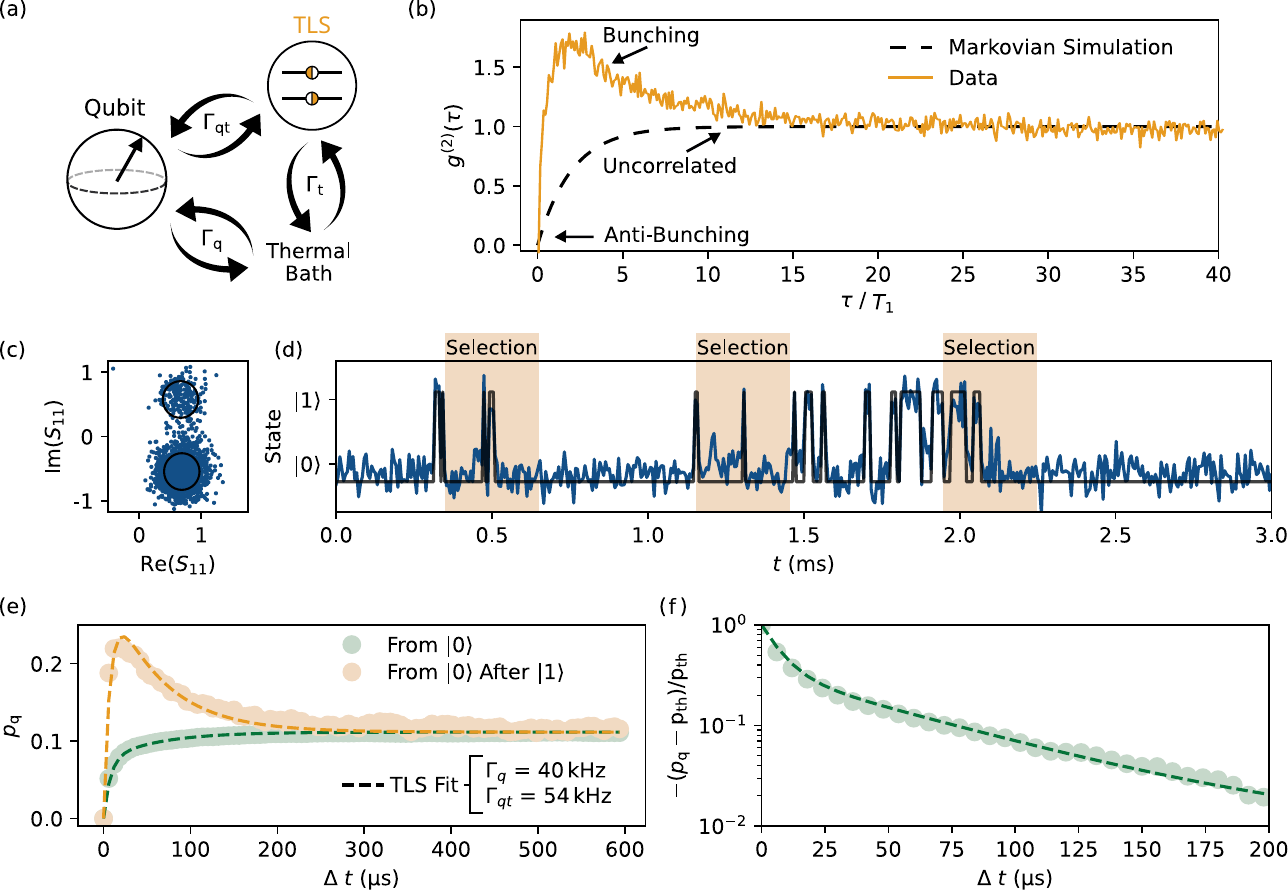}
\caption{\textbf{Non-Markovian qubit dynamics induced by coupling to a long-lived TLS.} \textbf{(a)} Schematics of the coupling of the qubit to both a TLS and a thermal bath. The coupling rates of the qubit to the TLS and the thermal bath are $\mathrm{\Gamma_{qt}}$ and $\mathrm{\Gamma_{q}}$, respectively. When the coupling rate of the TLS to the thermal bath $\mathrm{\Gamma_{t}}$ is smaller or comparable to $\mathrm{\Gamma_{qt}}$, the TLS becomes polarizable and we refer to it as a long-lived TLS. \textbf{(b)} Quantum jumps correlations from Solomon equation based simulation (dashed line) and from measured data (orange). As expected for a single photon emitter within the Born-Markov approximation (black dashed line), the $\mathrm{g_2}$ correlation function (see Eq. \ref{equ:g2}) shows anti-bunching of quantum jumps at short time scales and tends exponentially to uncorrelated jumps. In contrast, in the presence of a long-lived TLS, the correlation function shows quantum jump bunching persisting longer than the $T\mathrm{_1}$ of the qubit (orange). \textbf{(c)} IQ quadrature of 5000 single-shot readouts of the fluxonium qubit, plotted in the complex plane. The radius of the circles indicate 2 standard deviations for the distribution of the points. \textbf{(d)} Representative and cropped sample of the I quadrature over a 3 ms time trace, with the corresponding extracted state indicated by the continuous line.
Orange shaded areas highlight three examples of 0.5~ms long time traces, post selected after an energy relaxation event and contributing to the trace averaging.
\textbf{(e,f)} Qubit dynamics shown as averaged trajectories conditioned on ground-state selection (green) and jump-down events (orange). The data are well described by fits to the single-TLS Solomon model (eq. \ref{equ:Solomon}). 
\textbf{(f)} Zoom-in and re-scaling on the averaged dynamics after the ground state selection scenario, showing deviations from the expected exponential decay. 
}
\label{fig:correlation_technique}
\end{figure*} %2

While advances in circuit design and materials have extended superconducting qubit lifetimes into the millisecond regime~\cite{riste2013, thorbeck2023, bland2025, somoroff2023}, improved quantum control and measurement fidelity have revealed new decoherence phenomena that elude characterization by conventional $T\mathrm{_1}$ energy relaxation times.
Measurements of non exponential qubit decay~\cite{Pop2014, Gustavsson2016}, non-Poissonian quantum jumps~\cite{UriVool, SpieckerSolomon}, and hyperpolarization of the qubit environment~\cite{SpieckerTLSHyper, Odeh2025, zhuang2025}, point to the existence of a memory effect in qubit dissipation.
This violates the Markov approximation and underscores why $T\mathrm{_1}$ alone is an insufficient metric.
An immediate practical consequence for error correction algorithms is the appearance of time correlations in error syndromes, for example, due to a syndrome qubit flagging the same error multiple times. 
Note that these temporal correlations are different from the spacial correlation of several qubits on the same chip experiencing bit-flip errors simultaneously due to ionizing radiation impacts in the device substrate~\cite{Swenson2010, vepsalainenimpact, Cardani2021, Wilen2021, thorbeck2023, googleqec_corr, valenti2025}.

The existence of a memory in the qubit environment has recently been associated with so-called long-lived two-level systems (TLSs) that exceed by orders of magnitude the qubit lifetime~\cite{SpieckerTLSHyper, Liu2024Oct, Odeh2025}. 
In this work, we leverage the temporal correlation in the qubit's quantum jump dynamics to quantify and disentangle the coupling to long-lived TLSs from a Markovian background.
In contrast to previous pump–probe experiments \cite{SpieckerTLSHyper, zhuang2025}, this method involves no qubit manipulation. 
Moreover, continuous monitoring of the qubit via quantum jumps eliminates the need for reset times between measurement cycles.
Using quantum jump $correlation$ $spectroscopy$ at different qubit frequencies and electric field biases we uncover a resolved spectrum of electric field susceptible TLS memories, on top of a mostly featureless Markovian environment.

\begin{figure}[ht!]
\centering
\includegraphics[width=0.48\textwidth]{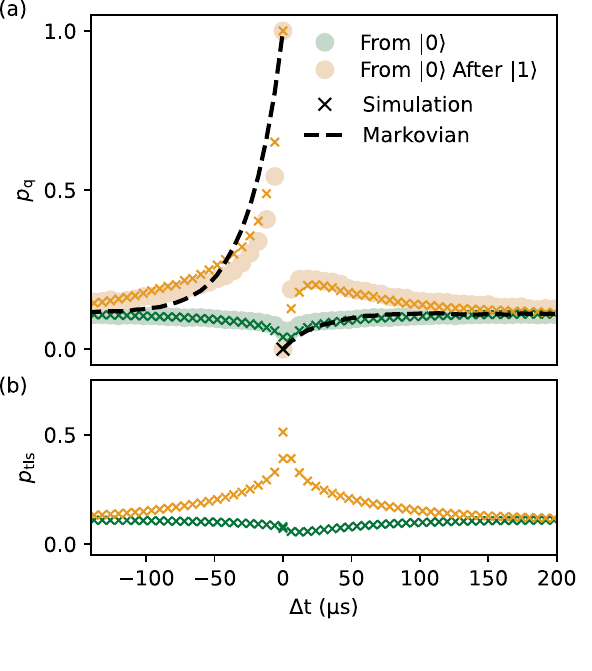}
\caption{
\textbf{Unraveling of the hidden TLS dynamics.}
Qubit (a) and hidden TLS (b) populations for the qubit at $\mathrm{\Delta}t$ = 0 in the ground state (green) or immediately after a jump down event (orange). The filled markers show measurements. Crosses show populations calculated from quantum jumps simulations, using the $\mathrm{\Gamma_{qt}}$, $\mathrm{\Gamma_{q}}$ and $\mathrm{\Gamma_{t}}$ extracted from the Solomon equations fit of the measured $\mathrm{P_q}$ for $\mathrm{\Delta t}$ $>$ 0. The dashed line shows populations from quantum jumps simulations assuming a Born-Markov environment. 
}
\label{fig:SimFit}
\end{figure}
 %3
\begin{figure*}[ht!]
\centering
\includegraphics[width=1\textwidth]{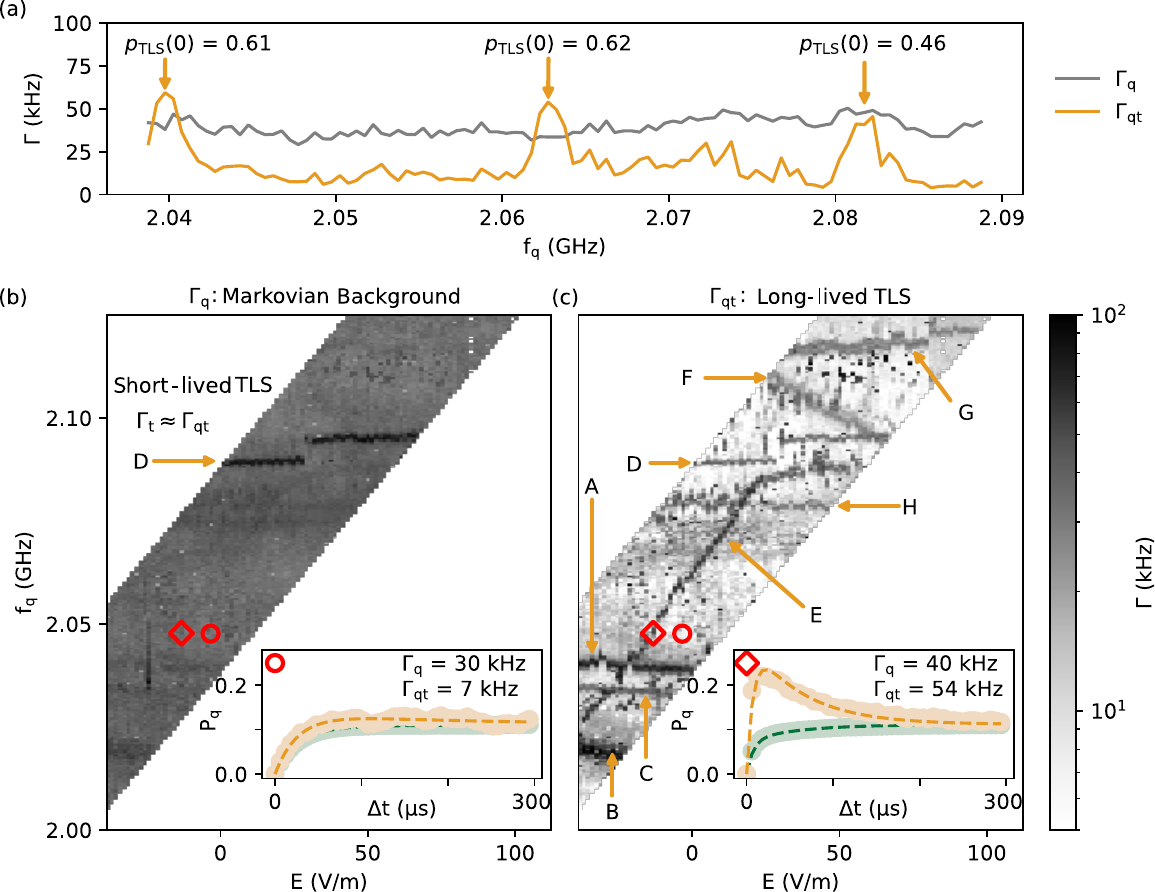}
\caption{
%\textbf{Correlation spectroscopy vs electric field bias}. 
\textbf{Electric field susceptibility of the long-lived TLSs}. 
(a) Single electric field correlation spectroscopy vs qubit frequency at 0 V/m. The Markovian environment of the qubit is nearly featureless, as shown by the qubit relaxation rate $\Gamma_q$ (b), while the qubit-TLS decay rate $\Gamma_{qt}$ shows distinct peaks susceptible to electric field bias (c). Insets show the qubit dynamics at the same frequency $f_q = \SI{2.045}{GHz}$ preformed under two different electric fields, corresponding  to a mostly Markovian environment ($\circ$) and a long-lived TLS in resonance ($\diamond$), respectively. To aid the discussion in the main text the long-lived TLSs are labeled A-H.}
\label{fig:spectro_vs_voltage}
\end{figure*} %4

A defining feature of a quantum two-level system is that after emitting a photon, it cannot emit another until it is re-excited. 
In the case of a continuously monitored qubit this implies an antibunching of quantum jumps. 
Yet, while superconducting qubits have been shown to behave as emitters of uncorrelated single-photons~\cite{lang2011g2,bozyigit2011g2}, the presence of even a single long-lived TLS (see Fig.~\ref{fig:correlation_technique}a) in their environment gives rise to bunching of quantum jumps~\cite{spiecker2024PhD}.
To quantify this we use the $\mathrm{g_2}$ correlation function:
\begin{equation}
\label{equ:g2}
g_2 = \frac{\sum_t P(\downarrow, t|\downarrow, t-\tau)}{P(\downarrow)},
\end{equation}
where $P(\downarrow, t|\downarrow, t-\tau)$ denotes the conditional probability of a jump down at a time t, given a jump down at time $t -\tau$, and $P(\downarrow)$ is the unconditional decay probability. For an ideal single-photon emitter, $g_{2}(\tau)$ starts at zero and exponentially approaches unity on the energy relaxation timescale $T_1$ (dashed line in Fig. \ref{fig:correlation_technique}(b)).

Experimentally, to extract $g_{2}(\tau)$ for a superconducting qubit, we obtain a quantum jump trace by performing stroboscopic readout every 6 $\mathrm{\upmu s }$ (cf.~Fig.~\ref{fig:correlation_technique}(c,d)), identify all downward jump events, average the subsequent time traces and normalize them by $P(\downarrow)$.
Figure~\ref{fig:correlation_technique}(b) shows results obtained on a fluxonium qubit (For information on the device see Appendix~\ref{App:qubit_presentation}), biased at $f_q = 2\,\mathrm{GHz}$ near the half-flux sweet spot. Although we observe the expected antibunching at short $\tau$, for $\tau$ exceeding the qubit relaxation time $T_1$ we instead find bunching: the photon emission probability exceeds its steady-state value.

The bunching of quantum jumps over times scales much longer than the qubit's $T\mathrm{_1}$ is indicative of the presence of a long-lived TLS in its environment ~\cite{SpieckerTLSHyper, Odeh2025, zhuang2025}. 
These long timescales also exclude an influence of the population of the qubit's higher levels, possibly induced by the readout~\cite{Dumas2024Oct,Dai2026Jan,Fechant2025Oct}. 
To describe the interaction between the qubit, a TLS and the thermal bath, we define key interaction rates: $\mathrm{\mathrm{\Gamma_q}}$ and $\mathrm{\Gamma_t}$, which describe the decay of the qubit and TLS, respectively, into a thermal bath, and $\mathrm{\Gamma_{qt}}$, the coupling rate governing energy exchange between the qubit and the TLS (cf.~Fig.\ref{fig:correlation_technique}(a)).
When $\mathrm{\Gamma_t} \lesssim \mathrm{\Gamma_{qt}}$ the TLS becomes polarizable and we refer to it as a long-lived TLS.
The dynamics of the coupled qubit-TLS system are described by the Solomon equations \cite{Solomon1955, SpieckerSolomon} (see Appendix \ref{App:SolMod}), in which joint evolutions of excitation probabilities $p_\mathrm{q}$ for the qubit and $p_\mathrm{t}$ for the TLS are
\begin{equation}
\label{equ:Solomon}
\frac{d}{dt}
\begin{pmatrix}
{p}_\mathrm{q} \\
{p}_\mathrm{t}
\end{pmatrix} = -
\begin{pmatrix}
\mathrm{\Gamma_q} + \mathrm{\Gamma_{qt}} & \mathrm{-\Gamma_{qt}} \\
-\mathrm{\Gamma_{qt}} & \mathrm{\Gamma_{qt}} + \mathrm{\Gamma_t}
\end{pmatrix}
\begin{pmatrix}
{p}_\mathrm{{q}} \\
{p}_\mathrm{{t}}
\end{pmatrix}
+ 
\begin{pmatrix}
\mathrm{\Gamma_q} \\
\mathrm{\Gamma_t}
\end{pmatrix}
p_\mathrm{{th}},
\end{equation}
where $p_\mathrm{{th}}$ denotes the thermal population for both the qubit and the TLS. 

Let's consider the dynamics immediately after the qubit is measured in the ground state, where only excitations are possible: $\dot{p}_\mathrm{q}(t)\big|_{p_\mathrm{q}=0}=\mathrm{\Gamma_\uparrow}(t)$. According to Eq.~\ref{equ:Solomon}, the excitation rate of the qubit depends on the population of the long-lived TLS coupled to it, $\mathrm{\Gamma_\uparrow(t)=\Gamma_{qt}} p_\mathrm{t}\mathrm{(t)}+\mathrm{\Gamma_q} p_\mathrm{{th}}$. 
If the TLS is approximately in thermal equilibrium with the environment, $p\mathrm{_t(t)}=p_{\mathrm{th}}$, its average effect is simply to increase the qubit relaxation rate by $\mathrm{\Gamma_{qt}}$.
This behavior is illustrated by the measured monotonic relaxation of the qubit from the ground state to $p\mathrm{_{th}\approx0.1}$ (green points in Fig.~\ref{fig:correlation_technique}(e)), even though small deviations from a purely exponential decay remain visible on a log--lin scale (see Fig.~\ref{fig:correlation_technique}(f)).
In contrast, consider trajectories immediately following a quantum jump from the excited to the ground state. 
In this case the TLS population is above its thermal value, as the TLS has a significant chance of having absorbed the photon emitted by the qubit.
This excess population temporarily increases the qubit excitation rate, resulting in non-monotonic and non-Markovian dynamics (orange points in Fig.~\ref{fig:correlation_technique}(e)) and explains the observed photon bunching.
Moreover, if qubit and TLS are each other's dominant relaxation channels ($\mathrm{\Gamma_{qt}>\Gamma_q}$), this energy exchange can repeat several times before the qubit excitation is ultimately released into the bath.

To extract the coupling rates defined in Fig.~\ref{fig:correlation_technique}(a), we fit the qubit relaxation conditioned on its previously measured state using the one-TLS Solomon model (Eq.~\ref{equ:Solomon}) (cf. Appendix \ref{app:2TLS} for a fit with more than one TLS).
We consider two simple ground to thermal-state relaxation scenarios: $\{\ket{0}\}$ corresponding to unconditional relaxation, and $\{\ket{1},\ket{0}\}$ corresponding to relaxation immediately following a downward jump. More complex selection scenarios, with different qubit dynamics, are discussed in Appendix~\ref{App:manyselection}. 
The fit is performed jointly on data from both selection scenarios to extract $\Gamma_\mathrm{q}$ and $\Gamma_{\mathrm{qt}}$. 
When $\Gamma_{\mathrm{t}}$ is included as a fit parameter, the resulting rate is typically orders of magnitude smaller than $\Gamma_\mathrm{q}$ and $\Gamma_{\mathrm{qt}}$ (cf. Appendix~\ref{App:gtasfit}); we therefore set $\Gamma_{\mathrm{t}} = 0$.
For the $\{\ket{0}\}$ scenario, the initial TLS population is assumed to be thermal. 
For the $\{\ket{1},\ket{0}\}$ selection scenario, we approximate the initial TLS population as
\begin{equation}
\label{equ:ptls}
p_{\mathrm{t}}(0) = p_{\mathrm{th}} + (1 - p_{\mathrm{th}})\frac{\Gamma_{\mathrm{qt}}}{\Gamma_\mathrm{q} + \Gamma_{\mathrm{qt}}},
\end{equation}
where $\Gamma_{\mathrm{qt}}/(\Gamma_\mathrm{q} + \Gamma_{\mathrm{qt}})$ denotes the probability that the qubit decays into the TLS when the TLS is empty (with probability $1 - p_{\mathrm{th}}$).
As illustrated by the dashed lines in Fig.~\ref{fig:correlation_technique}(e), this model quantitatively describes the measured qubit dynamics.  
An immediate consequence is that the standard description using a single, constant decay rate $1/T\mathrm{_1}$ is no longer accurate. 
The qubit now has two decay rates, $\mathrm{\Gamma_q}$ and $\mathrm{\Gamma_{qt}}$. 
When needed, this analysis directly extends to a qubit simultaneously coupled to multiple TLSs, as discussed in Appendix~\ref{App:SolMod}, in which case we need to keep track of each qubit-TLS interaction rate. 

Each qubit measurement selection scenario prepares a specific, unobservable TLS population that can be unraveled~\cite{Gneiting2021Dec, SpieckerTLSHyper} from the measured quantum jump trace.
To uncover the hidden TLS populations, we perform a Monte-Carlo simulation of quantum jump traces (cf. Appendix~\ref{App:MCSim}), using the Solomon equations and the fitted relaxation rates. 
For the two simplest scenarios, $\{\ket{0}\}$ and $\{\ket{1},\ket{0}\}$ (cf. Fig.~\ref{fig:correlation_technique}(e,f)), the unraveled TLS and qubit dynamics are plotted in Fig.~\ref{fig:SimFit}.
In the $\{\ket{1},\ket{0}\}$ scenario, we uncover an increase of the hidden TLS population immediately before a jump down event, which decays back to the thermal state. 
Using these simulations, we verify that Equation~\ref{equ:ptls} provides a good approximation to the TLS population at the instant of the $\{\ket{1},\ket{0}\}$ selection scenario (cf. Appendix~\ref{App:ptcompare}).
This observation is in contrast with the ground state selection $\{\ket{0}\}$, in which case the TLS population decreases below equilibrium.

TLSs have been observed in superconducting qubits for more than two decades \cite{Barends2013Aug, Martinis2005Nov, Zagoskin2006Aug, Lisenfeld2016Mar, Liu2024Oct}, yet their physical origin remains unclear, including whether long-lived TLSs belong to the same family of defects as the broader TLS population.
To obtain the spectrum of the long-lived TLS and gain further insight into their nature, we perform spectroscopy by repeating the quantum jump traces with our fitting procedure versus qubit frequency. We refer to this method as quantum jump $correlation$ $spectroscopy$. 
In Fig.~\ref{fig:spectro_vs_voltage}(a), the fitted interaction rates over a 50~MHz span show individual peaks in the coupling rate $\mathrm{\Gamma_{qt}}$, above the relatively flat Markovian background (for a larger span of 500 MHz see appendix \ref{App:branches}) . 
Each of these peaks is indicative of the presence of a long-lived TLS in the qubit environment at the corresponding frequency.

Next, we perform the correlation spectroscopy while applying an electric field bias though a metallic pin positioned several millimeters from the qubit, to investigate the susceptibility of the long lived TLS environment to an electric field.
The fluxonium qubit spectrum is insensitive to charge bias \cite{Manucharyan2009Oct}, therefore we only expect changes in the environment that is electric field susceptible.
As shown in Fig~\ref{fig:spectro_vs_voltage}(b,c), the correlation spectroscopy reveals peaks related to long-lived TLS susceptible to electric field.
The spectroscopy tracks the frequency of a TLS, which increases with applied field labeled E in Figure~\ref{fig:spectro_vs_voltage}(c). We also encounter TLSs whose frequency decrease with field bias at different rates labeled B, C and F, stay constant labeled A, D, G and H, or jump in time labeled D and G. 
This phenomenology for long-lived TLSs is similar to the previously reported qubit environment spectroscopy, with the distinction that here we can disentangle short-lived from long-lived TLS species. 
Indeed, a short-lived TLS labeled D is visible in the Markovian background, indicating that its intrinsic relaxation is comparable to the interaction rate with the qubit (for details see Appendix~\ref{App:shortlived}).

%\lipsum[1-10]

In this letter, we have shown that quantum jumps of a superconducting qubit unveil long-lived memory effects in its environment, on timescales orders of magnitude longer than the qubit's intrinsic energy relaxation.
While these effects can go unnoticed in standard $T_1$ relaxation analysis, they have profound consequences for the dynamics of the underlying quantum system, introducing time-correlated errors and challenging the prevailing quantum error correction approaches.
By measuring quantum jumps spontaneously induced by thermal fluctuations in the qubit–TLS system and fitting the resulting conditional dynamics using the Solomon equations, we isolate the contributions of a conventional Markovian bath from those of individual, long-lived TLSs. 
We observe distinct resonant features in the qubit-TLS coupling rates by sweeping the qubit frequency, directly identifying long-lived TLSs across frequency spans of tens of megahertz within a few minutes.
Furthermore, we showed that long-lived TLSs can be susceptible to electric field.

The quantum jump correlation spectroscopy introduced here offers a ready-to-use tool for quantitatively mapping long-lived TLSs, even in the weak-coupling regime where they might otherwise remain undetected.
Our technique is deployable on any standard qubit measurement setup and does not require qubit manipulation, eliminating gate calibration errors or drive-induced leakage errors~\cite{Fechant2025Oct, Dai2026Jan}. 
Moreover, this technique bypasses the need for qubit initialization and feedback control by harnessing thermal fluctuations.
While in the present system a semi-classical rate-equation description is sufficient, more advanced methods that take into account possible entanglement between the qubit and the long-lived TLS could also be deployed~\cite{SpieckerSolomon}.
As control over superconducting qubit technology and materials improves, and error correction becomes integral to quantum devices, mitigating long-lived environmental memory effects will become an increasingly stringent necessity.

\section*{Data Availability}
All relevant data are available from the corresponding author upon reasonable request.

\section*{Acknowledgments}
We are grateful to Jürgen Lisenfeld, Andrei I. Pavlov, and Enzo Schmitz for stimulating discussions. 
We thank L. Radtke and S. Diewald for technical assistance.
This work was financed by the German Federal Ministry of Research, Technology and Space (BMFTR) within project QSolid (FKZ:13N16151).
D.B.~acknowledges funding from the Horizon Europe program via Project No.~101113946 OpenSuperQPlus100.
P.K.~acknowledges funding from the European Partnership on Metrology, co-financed from the European Union’s Horizon Europe Research and Innovation Programme and by the Participating States (contract number: 23FUN08 MetSuperQ).
Facilities use was supported by the KIT Nanostructure Service Laboratory. 
We acknowledge the measurement software framework qKit.
%\nocite{*}

\bibliography{references}

\balancecolsandclearpage

\onecolumngrid

\section*{Appendices}
\vspace{0.6cm}
\appendix

\section{Qubit presentation}
\label{App:qubit_presentation}

\begin{figure}[ht!]
\centering
\includegraphics[width=1\textwidth]{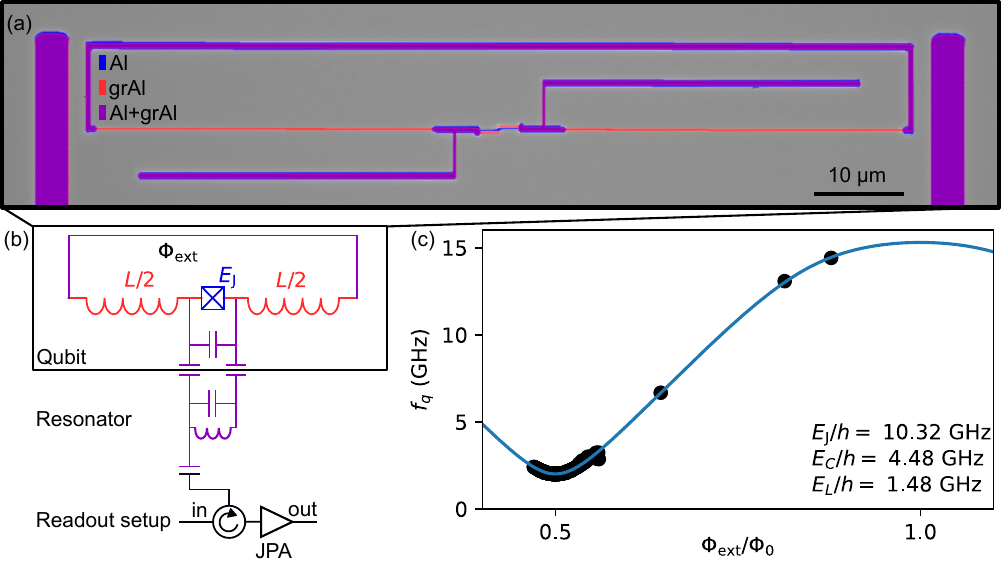}
\caption{
\textbf{Granular aluminum fluxonium qubit}. (a) False-colored optical microscope picture of the qubit sample. The stack is composed of two aluminum layers (blue) and one granular aluminum layer (red). (b) Equivalent circuit of the fluxonium qubit, capacitively coupled to an aluminum lumped-element LC resonator for readout. The resonator is measured in reflection via a circulator, and the signal is amplified by a low-noise Josephson Parametric Amplifier (JPA) \cite{winkel_nondegenerate_2020} (c) Qubit spectrum versus external flux $\Phi_\mathrm{ext}$. Black dots denote the measured $0-1$ transition frequencies, solid line depicts the fit to the fluxonium Hamiltonian. 
}
\label{fig:qubit_presentation}
\end{figure} 

The qubit is a fluxonium qubit where the inductance is realize with a granular aluminum wire, similarly to Refs. \cite{grunhaupt_granular_2019, denis}. 

The sample was fabricated on a single double-side polished c-plane 330-$\si{\micro m}$-thick sapphire wafer. A double stack of MMA(8.5)MAA EL13 and 950 PMMA A4 resists was spincoated on the wafer and patterned using a $\SI{50}{keV}$ e-beam lithography tool. After developing in a IPA/$\mathrm{H}_2$O 3:1 solution for $\SI{2}{min} \, \SI{30}{s}$, aluminum was deposited in a three-angle shadow evaporation process in a Plassys e-beam evaporation system. The thin film deposition recipe includes plasma cleaning and Ti gettering to clean the sample and the chamber, $\SI{-34}{\degree}$-tilted aluminum evaporation at $\SI{1}{nm/s}$ for \SI{13}{s}, 6-$\si{min}$ oxidation at $\SI{10.1}{mbar}$, $\SI{34}{\degree}$-tilted aluminum evaporation at $\SI{1}{nm/s}$ for \SI{32}{s}, argon milling to ensure contact, and granular aluminum deposition consisting of a $\SI{0}{\degree}$ aluminum evaporation at $\SI{1}{nm/s}$ for \SI{55}{s} under a constant oxygen flow at $\SI{6.7}{sccm}$. After lift-off in hot acetone ($\SI{50}{\degree C}$, $\SI{16}{h}$) and hot DMSO ($\SI{90}{\degree C}$, $\SI{20}{min}$), the wafer was cleaned in acetone and IPA. Finally, the wafer was diced in $\SI{15}{mm}\,\times\,\SI{3}{mm}$ chips after coating it with a protective S1818 G2 photo-resist, which is removed in hot DMSO ($\SI{90}{\degree C}$, $\SI{5}{min}$). 

In Fig. \ref{fig:qubit_presentation}a, we show a picture of the sample in false colors, with aluminum in blue, forming the junction in the middle, granular aluminum in red that creates the inductive wires and a stack of aluminum and granular aluminum making up the rest of the structures. The qubit couples to an aluminum lumped-element LC resonator whose capacitive pads are visible on the right and left hand sides. Figure \ref{fig:qubit_presentation}b depicts the equivalent circuit of the qubit and resonator system. To read the qubit state, we measure the resonator frequency in reflection, in a standard dispersive readout scheme. The signal is amplified at base temperature by a Josephson parametric amplifier made in-house \cite{winkel_nondegenerate_2020}. Finally, we measure the qubit frequency in a two-tone spectroscopy and fit it to the fluxonium Hamiltonian
\begin{equation}
    \hat{\mathcal{H}} = 4 E_{C} \hat n^2 + \frac{E_L}{2} (\hat\varphi+\varphi_\mathrm{ext})^2 - E_\mathrm{J} \cos\hat\varphi,
\end{equation}
and we obtain the following fluxonium parameters $E_\mathrm{J}/h=\SI{10.32}{GHz}$, $E_C/h=\SI{4.48}{GHz}$, $E_L/h=\SI{1.48}{GHz}$, as depicted in Fig. \ref{fig:qubit_presentation}c.

\newpage

\section{Solomon Model}

Following~\cite{SpieckerSolomon}, the Solomon model can easily be expanded to multiple TLSs by increasing the size of the interaction matrix $\mathrm{\textbf{M}}$ and of the vector of populations $\mathrm{\textbf{p}}$ : 
\label{App:SolMod}

\begin{equation}
\label{equ:Solomon_supplementary}
\mathbf{\dot{p}} = -
\mathbf{M}
\mathbf{p}
+ 
\begin{pmatrix}
\Gamma_q \\
\mathbf{\Gamma_t} 
\end{pmatrix}
\mathrm{p_{th}}
\end{equation}

\begin{equation}
\mathbf{M} = 
\begin{pmatrix}
\Gamma_q + \sum_i \Gamma_{qt}^i & -\mathbf{\Gamma_{qt}} \\
-\mathbf{\Gamma_{qt}} & \mathbf{\Gamma_{tls}}
\end{pmatrix}
\end{equation}
Where $\Gamma_q$ denotes the decay of the qubit into a thermodynamic environment. $\mathbf{\Gamma_{qt}}$ is a vector characterizing the rates between the qubit and each TLS. $\mathbf{\Gamma_{tls}}$ is a diagonal matrix where the entries correspond to the sum of the individual coupling rates $\Gamma_{qt}^i$ of the i'th TLS and the qubit, and the TLS decay to the thermal bath $\Gamma_t$.
\begin{equation}
\mathbf{\Gamma_{TLS}} = 
\begin{pmatrix}
\Gamma_{qt}^1 + \Gamma_{t} & 0 & ... & 0\\
0 & \Gamma_{qt}^2 + \Gamma_{t} & ... & 0 \\
\vdots  & \vdots  & \ddots  & \vdots  \\
0 & ... & 0 &  \Gamma_{qt}^n + \Gamma_{t}\\
\end{pmatrix}
\end{equation}

\newpage

\section{2 TLS Fit}
\label{app:2TLS}

In Figure \ref{fig:spectro_vs_voltage} you can see Points at which the coupling to a single TLS is strong compared to the thermal background (such as at 2.05 GHz and -10 V/m). However there are also points in between two strong coupling such as where the TLS A and E meet. Those points require more than a single TLS in the model to completely predict the qubit dynamics especially for longer times.

\begin{figure}[ht!]
\includegraphics[width=1\textwidth]{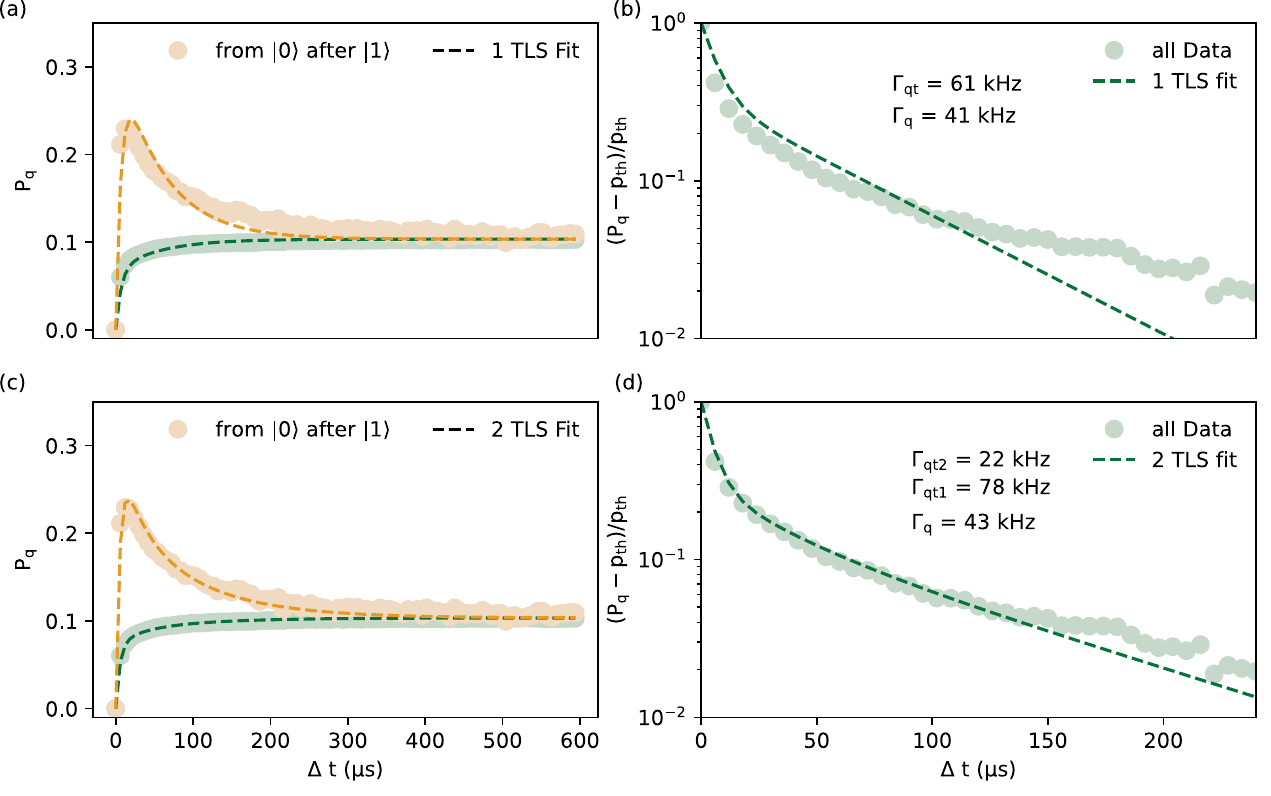}
\caption{\textbf{Fitting with more than one TLS} (a) and (b) show a fit using the Solomon equations from the main text. (c) and (d) show a fit using 2 TLSs coupled to the qubit. }
\label{fig:App2TLSFit}
\end{figure} %3

The Solomon equations used for the fitting routine naturally expand to multiple TLSs. In the 2 TLS case we now need to coupling rates $\mathrm{\Gamma_{qt}^1}$ and $\mathrm{\Gamma_{qt}^2}$.  The qubit decay rate $\mathrm{\Gamma_{q}}$ and the thermal population $\mathrm{p_{th}}$ are still single values and $\mathrm{\Gamma_{t}}$ is also set to 0 as in previous fits.

\newpage
\section{More Complex Selection Scenarios}
\label{App:manyselection}

\begin{figure*}[h]
\centering
\includegraphics[width=\textwidth]{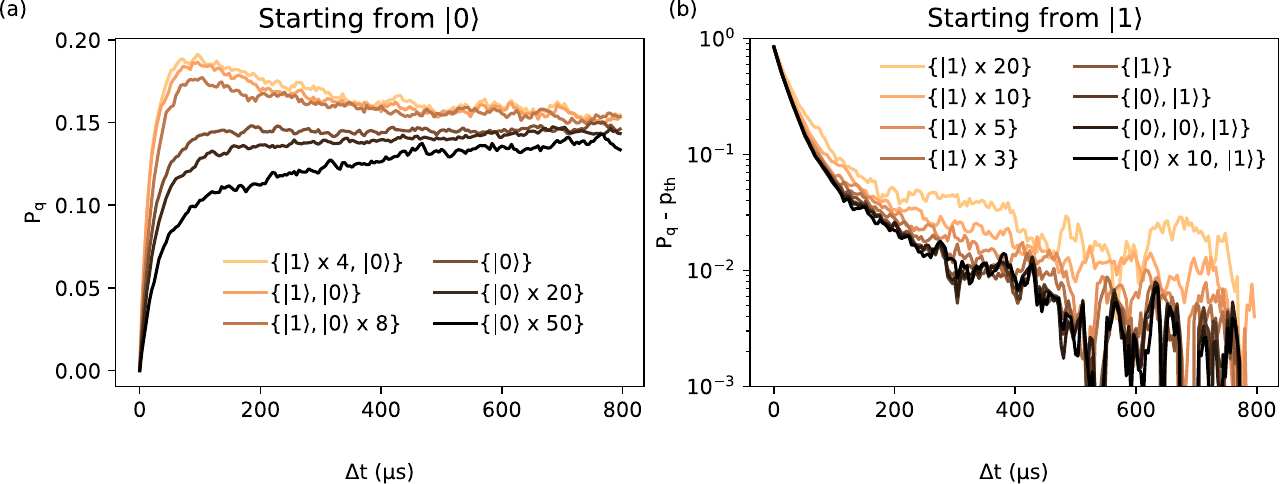}
\caption{
\textbf{More complex selection scenarios}. Measured qubit dynamics starting from the ground state (a) and the excited state (b) for various selection scenarios. The convention used for the labeling is that \{\} indicate and ordered list and $\ket{q}$ x N indicates the state $q$ repeats N times.
}
\label{fig:manyselection}
\end{figure*}

For simplicity, in the main text Fig.~\ref{fig:correlation_technique} we show the dynamics of the qubit resulting from two selection rules $\{\ket{0}\}$ and $\{\ket{1},\ket{0}\}$, which contain sufficient information to quantitatively fit the rates in the Solomon equations. 
For completeness, Fig.~\ref{fig:manyselection} shows additional measured qubit dynamics following more complex selection rules, starting from the ground state (a) and from the excited state (b).
Note that the more often the selection rule includes the qubit in the excited state, the higher the resulting population overshoot and the higher the underlying starting hidden TLS population.  

\newpage
\section{$\Gamma_\mathrm{t}$ as a Fit parameter}
\label{App:gtasfit}

\begin{figure*}[h]
\centering
\includegraphics[width=\textwidth]{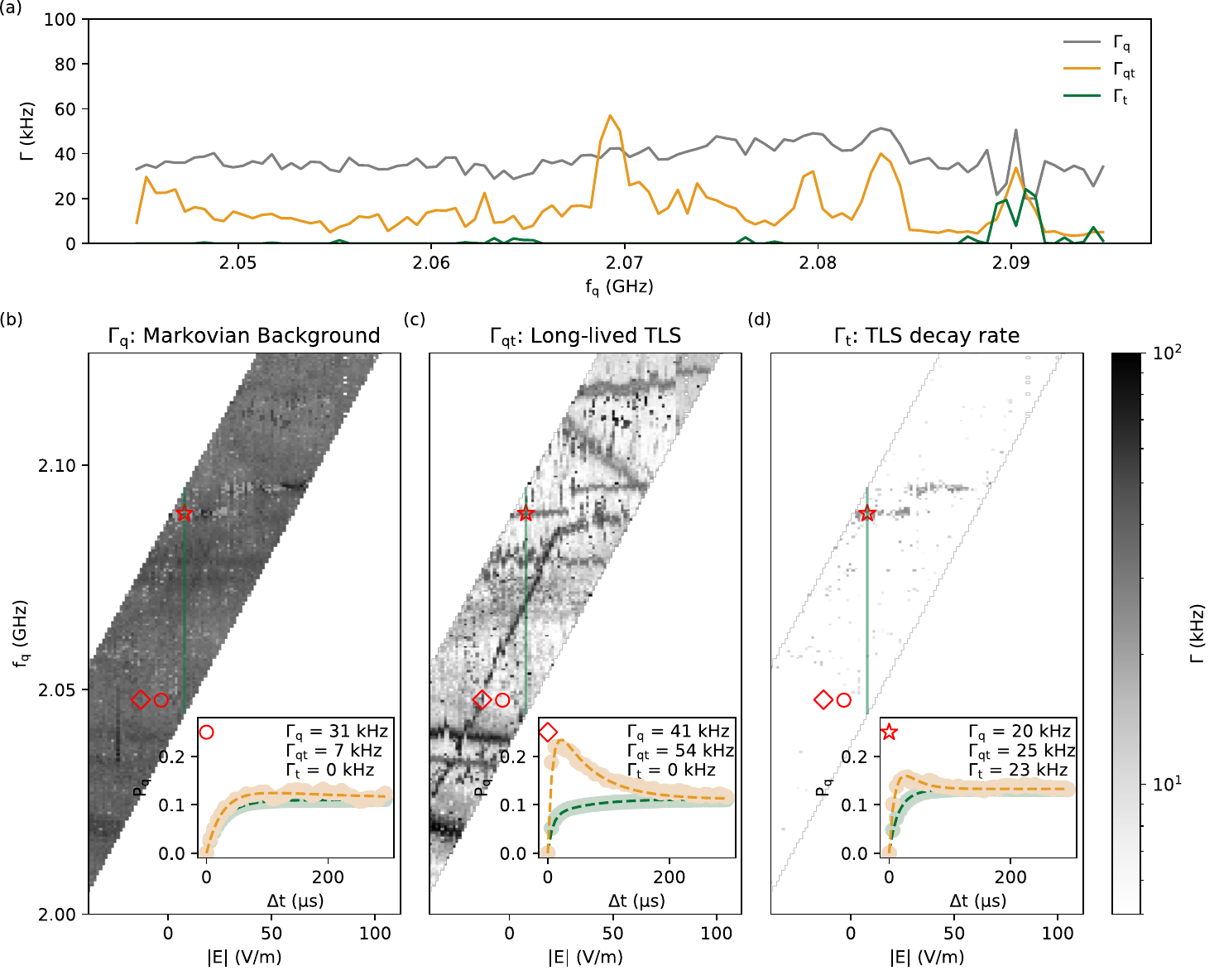}
\caption{
\textbf{Fitting of data from main text including $\Gamma_\mathrm{t}$ as a fit parameter}. (a) Interaction rates $\Gamma_\mathrm{t}$ (green), $\Gamma_\mathrm{q}$ (gray) and $\Gamma_\mathrm{qt}$ (orange) for a single trace fit at applied 8 $\mathrm{V/m}$ Electric field. (b) $\Gamma_\mathrm{q}$, (c) $\Gamma_\mathrm{qt}$ and (d) $\Gamma_\mathrm{t}$ for the full data set. Insets show selected qubit dynamics and fit at marked points.
}
\label{fig:gtasfit}
\end{figure*}

For simplicity, in the main text we do not fit the TLSs decay rate $\Gamma_\mathrm{t}$. Here we show that including it as a fit value, does not improve the quality of the fit. However it does mostly remove the signature of the short lived TLS from the Markovian environment (vf.Fig~\ref{fig:gtasfit}). Moreover for most cases this fit shows that setting the TLSs decay rate to 0 is sufficient for any long-lived object.

\newpage
\section{Monte Carlo Simulation of Quantum Jumps}
\label{App:MCSim}

The simulations used in Fig.~\ref{fig:SimFit} of the main text employ the Monte Carlo wavefunction method~\cite{molmer1992}, which models quantum processes using stochastic quantum jumps. The procedure begins by evolving the state probabilities of the qubit and TLS for a short time interval ($\ll T_1$). A random number $\epsilon \in (0,1)$ is then drawn to determine a projection of the qubit state. The resulting state is obtained by comparing $\epsilon$ with the qubit excited-state population $p_q$ and the thermal equilibrium probability $p_{\mathrm{th}}$. If $p_q < \epsilon + p_{\mathrm{th}}$, the qubit is projected to the ground state. Otherwise the qubit is projected to the excited state.

Furthermore this method enables us to access the TLS population at any time step. This population is not measurable in experiment, however analyzing the effect of the post selection on the TLS population can give physical insight into the behavior of the qubit dynamics. This also allows us to double check the assumption that the population of the TLS after the jump down corresponds to: $P_{tls}(\downarrow) = \frac{\Gamma_{qt}}{\Gamma_q + \Gamma_{qt}}(1-p_{th}) + p_{th}$ (cf.~Appendix~\ref{App:ptcompare}). Indeed, both the analytical formula and the simulation give a starting TLS population of $p_{\mathrm{t}}(0) = 0.55$ for the $\{\ket{1},\ket{0}\}$ selection scenario. 

\newpage
\section{Compare $p_\mathrm{TLS}(0)$ from Simulation and Formula}
\label{App:ptcompare}

\begin{figure*}[ht!]
\centering
\includegraphics[width=0.5\textwidth]{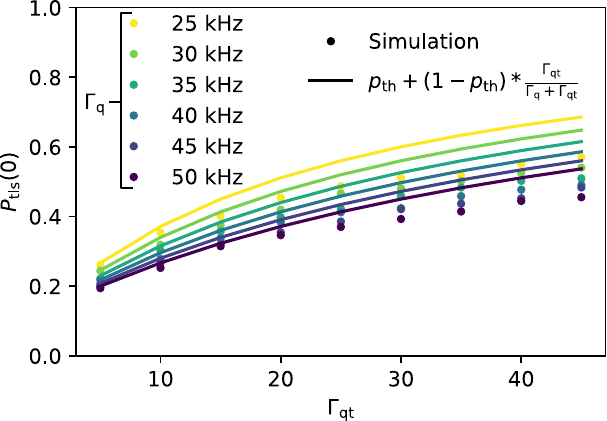}
\caption{
\textbf{Comparing the population of the TLS just after a jump down of the qubit from Simulation and Formula}. Simulation (points) and formula (line) of the population of the TLS after the qubit just jumped down, for different rates. 
}
\label{fig:Pt_compare}
\end{figure*}

In the main text we use Eq.~\ref{equ:ptls} to estimate the TLS population after a jump down. In figure \ref{fig:Pt_compare} we present a comparison between the formula and the TLS population extraced from a monte carlo simulation. The rates used in the simulation are sampled from the fitted values from the data (c.f. Fig.~\ref{fig:spectro_vs_voltage} in main text) in the main text.

\newpage

\section{Scanning TLSs multiple times}
\label{App:branches}

\begin{figure}[ht!]
\includegraphics[width=1\textwidth]{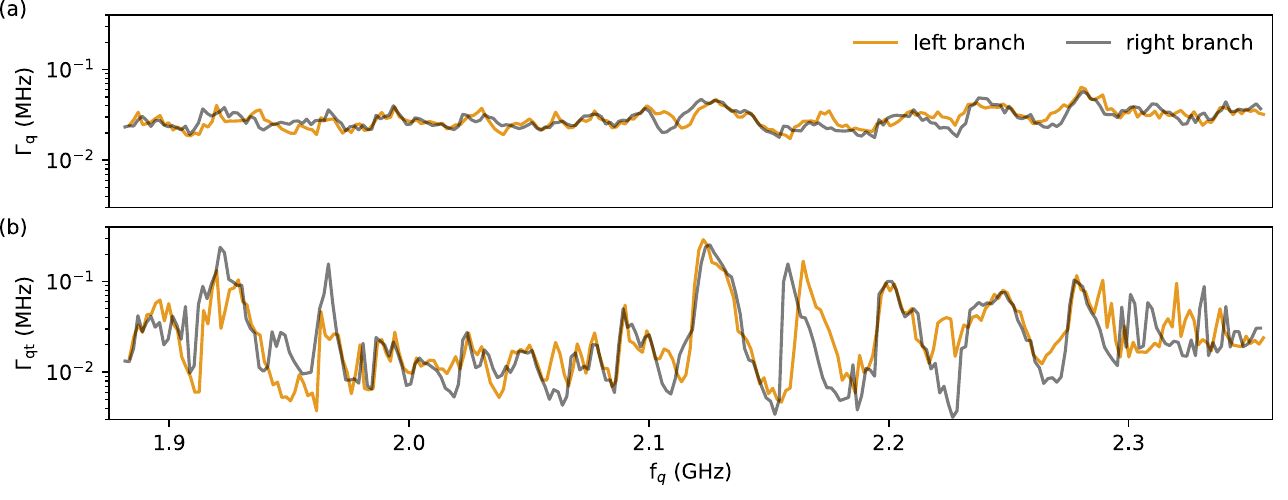}
\caption{\textbf{Correlation spectroscopy as a probe of long-lived TLSs over multiple branches around the qubit's half flux sweet spot.}.
}
\label{figapp:BranchTLS}
\end{figure}

In Figure \ref{figapp:BranchTLS} we show a larger spectroscopy for a similar qubit at 0 electric field. Here the orange and black lines correspond to the left and right side respectively of the fluxoniums spectroscopy centered around half flux. The data is taken with linear spacing in frequency beginning from the highest frequency in the right branch and ending at the highest frequency in the left branch. The furthest data points are taken 36 minutes apart. Most of the TLS stay at the same frequency through the experiment. The small visible shifts could be due to small changes in the environment or reshuffling events \cite{Swenson2010, vepsalainenimpact, Cardani2021, Wilen2021, thorbeck2023, googleqec_corr, valenti2025} happening during the measurement.

 %3
\newpage
\section{Short Lived TLS}
\label{App:shortlived}

\begin{figure*}[ht!]
\centering
\includegraphics[width=1\textwidth]{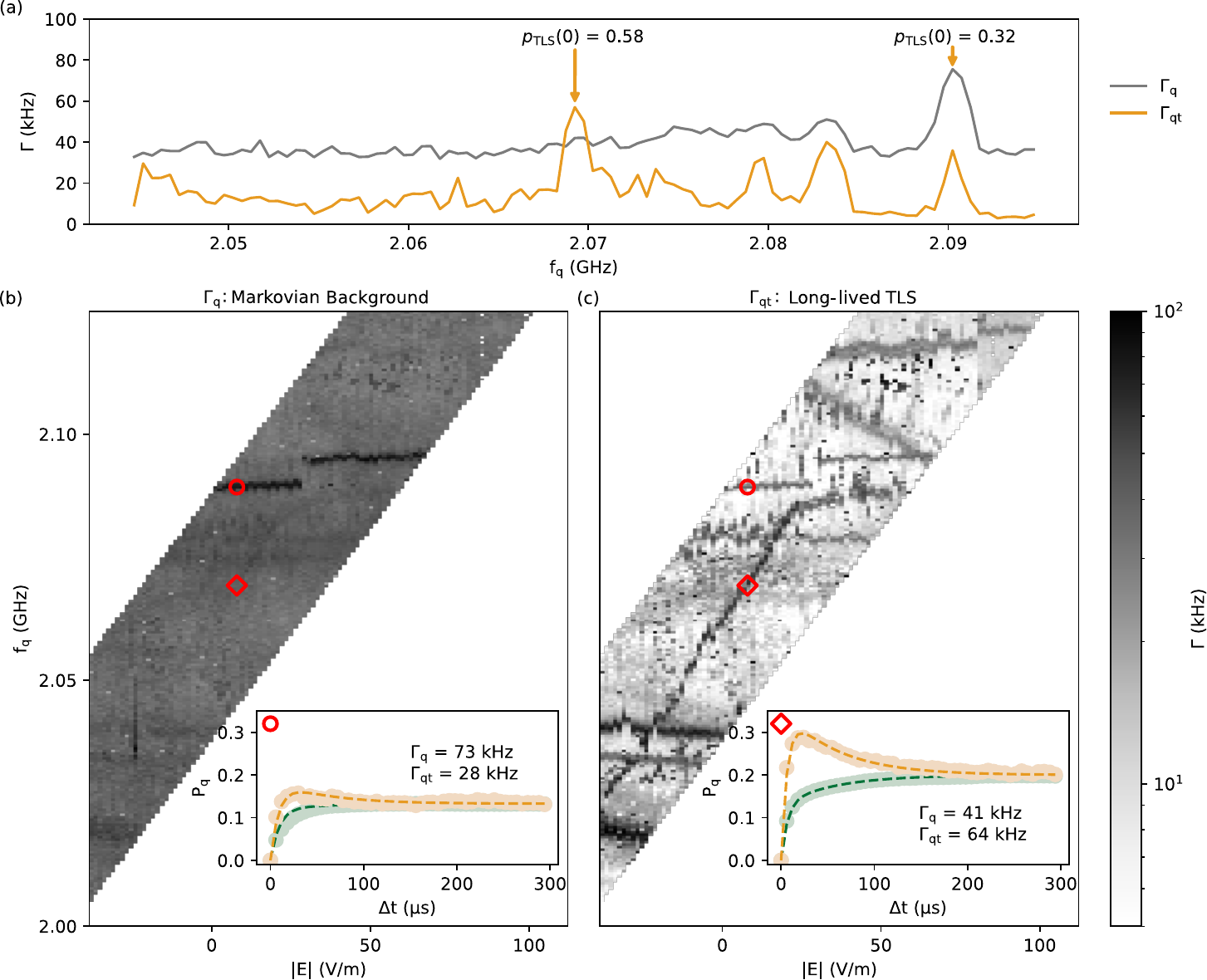}
\caption{
\textbf{Correlation spectroscopy vs electric field bias}. (a) Single electric field correlation spectroscopy vs qubit frequency at -10 v/m. The Markovian environment of the qubit is nearly featureless, as shown by the qubit relaxation rate $\Gamma_q$ (b), while the qubit-TLS decay rate $\Gamma_{qt}$ shows distinct peaks susceptible to electric field bias (c). Insets show the qubit dynamics for a short lived TLS in the Markovian background and a few kHz below it.}
\label{fig:shortspectro_vs_voltage}
\end{figure*}

In the main text in Fig.~\ref{fig:spectro_vs_voltage}(a) we show that in the Markovian background ($\mathrm{\Gamma_q}$) a TLS line is visible. We identify it as a short lived TLS and in Fig.~\ref{fig:shortspectro_vs_voltage} we show the fitted corelation spectroscopy around this short lived TLS. 
The fitting procedure holds $\Gamma_{\mathrm{t}}$ at 0 which in-turn makes losses from the TLS appear as losses from the qubit and therefore increases the $\Gamma_{\mathrm{q}}$ in the fit.

\end{document}